# Efficient inverted HTL-free Sm$_2$NiMnO$_6$-based perovskite solar cell: a SCAPS-1D study.

Authors: **Nassim Ahmed Mahammedi** [1, *]

[1]Department of physics, Stockholm University, 106 91 Stockholm, SWEDEN

*Corresponding Author: N. A. Mahammedi: nassim.mahammedi@fysik.su.se

## Abstract

The transition to sustainable energy has accelerated research into perovskite solar cells (PSCs) as promising candidates for next-generation photovoltaics. Despite their remarkable efficiencies, the commercialization of PSCs is hindered by lead toxicity and material instability. In this study, we explore a lead-free Samarium-based double perovskite oxide Sm$_2$NiMnO$_6$ (SNMO), as the active absorber layer in an innovative inverted HTL-free PSC architecture. Using SCAPS-1D simulations, we optimized the device configuration to achieve a power conversion efficiency (PCE) of 10.93% with an open-circuit voltage ($V_{OC}$) of 0.8 V, a short-circuit current density ($J_{SC}$) of 16.46 mA/cm², and a fill factor (FF) of 82.14%. We demonstrate that increasing the SNMO absorber thickness enhances light harvesting in the red spectrum region (~620 nm), shifting the external quantum efficiency (EQE) peak from 380 nm at 50 nm thickness to about 620 nm at 1 µm. We also investigated various electron transport layers (ETLs) and found that the indium tin oxide (ITO) exhibited superior PV performances, boosting the PCE to ~12.6% due to its excellent conductivity and optimal energy band alignment with SNMO. These findings establish SNMO as a promising absorber material for environmentally friendly PSCs, paving the way for cheaper, simpler, scalable, and sustainable photovoltaic solutions. This work highlights the potential of HTL-free architectures in reducing costs and complexities while maintaining competitive efficiencies, offering a significant step forward in advancing lead-free solar technologies.

## Keywords

## I. Introduction

Over the past two decades, Perovskite solar cells (PSCs) become one of the most investigated branches in the photovoltaic (PV) realm. A large research community is devoting tremendous theoretical and experimental efforts in order to unveil the hidden potentials of Perovskites as they are already revolutionizing the field, owing to their astonishing physical-chemical properties providing not only exceptional light absorption, tunable bandgaps and compatibility with low-temperature fabrication processes, but also a great ability to be easily integrated into the existing technologies (Si, a-Si, CIGS, CdTe, and other platforms) [1-3]. This versatility made perovskite the most attractive topic. Among the most promising perovskites, the rare-earth-based double perovskite oxides family (DPO) A$_2$BB'O$_6$ is gaining considerable attention; here, A is a rare-earth element, B and B' are transition metals. This eco-friendly multifunctional class of perovskites exhibited excellent properties such as adequate direct bandgaps, low-temperature ferromagnetism, large dielectric constant, higher magnetoresistance, and magnetocapacitance [4-7]. When Nickel and Magnesium are involved (X$_2$NiMnO$_6$), due to Ni-O-Mn electronic interaction, ferromagnetism dielectric and magnetoresistance properties are improved, making the X$_2$NiMnO$_6$ group suitable for various applications [7, 8]. Samarium-based double perovskite oxide is progressively rising as a promising material with an excellent bandgap (1.3 – 1.4eV) and a large dielectric constant. Many routes are explored in order to synthesize SNMO perovskite in various forms; *W. Z. Yang et al.* prepared $Ln_2NiMnO_6$ (Ln=Nd and Sm) perovskite ceramics using a solid-state sintering process [6]. Their results revealed a very large dielectric constant for SNMO at room temperature, ranging from 1250 for 10 Hz external electric field frequency to about 250 for $10^5$ Hz [6]. *Md Sariful Sheikh et al*. prepared SNMO nanoparticle nanoparticles using the sol-gel method. They reported a promising direct bandgap Eg ranging from 1.2eV to 1.41eV, making it an excellent choice for photovoltaics [8, 9]. They showed that each Ni and Mn is surrounded by the six O-ions, creating $NiO_6$ and $MnO_6$ octahedra, respectively. Therefore, the physical properties, such as electronic, optical, and magnetic properties of the SNMO are affected by various kinds of octahedral distortions, octahedron framework tilting, off-center displacement of the central atom of the octahedron, etc., as well as the size and the charge of the cations which are important to fine-tune the physical properties of SNMO by these octahedral distortions. Moreover, SNMO demonstrated a significant carrier lifetime of ~ 10.2 µs, high photocurrent density (in mA range), and higher light absorption coefficient in the visible region reaching up to ~ $10^5$ cm$^{-1}$[8], which is comparable to that of Si ($\alpha$~$10^4$ cm$^{-1}$) and higher to that of GaAs ($\alpha$~$10^4$ cm$^{-1}$) [8]. *R.J. Booth et al.* prepared several perovskite oxides for the composition $R_2NiMnO_6$ (R = La, Pr, Nd, Sm, Gd, Tb, Dy, Ho, and Y) by reacting high purity $R_2O_3$ (pre-dried at 900 °C), $NiC_2O_4.2H_2O$ and $MnC_2O_4.2H_2O$ in stoichiometric proportion at different synthesis conditions [10]. *Kai Leng et al* prepared SNMO nanoparticles using the sol-gel technique; they reported a direct optical band gap of 1.42 eV, close to the Shockley-Queisser band gap (1.34 eV) of a single junction solar cell with solar cell efficiency over 33% proving that SNMO nanoparticles could be useful for PV devices[11]. *Felicia Gheorghiu et al.* prepared SNMO ceramics by spark plasma sintering (SPS) from powders synthesized by sol-gel self-combustion, in this research, SNMO ceramics with high crystallinity levels corresponding to the monoclinic structures were reported. A non-linear dielectric character was reported for the first time in SNMO, which means that the dielectric constant depends on the external electric field frequency and the synthesis method. Due to its multiferroic character, they also obtained strong nonlinearity and small hysteretic behavior, making the SNMO system a promising candidate for different modern applications [12]. PSCs, including the Ln2NiMnO6-based SCs, usually consist of a perovskite absorbing layer sandwiched between a hole transport layer (HTL) and an electron transport layer (ETL), depending on the cell arrangement; the solar cell could be either in an n-i-p or p-i-n configuration [3, 13]. There are very few studies on SNMO-based solar cells in the literature. However, most of the studies are devoted to the La2NiMnO6 (LNMO) counterpart [5, 14, 15]. Solar cells based on LNMO showed promising performances, for example, *M. Khalid Hossain et al.* obtained $V_{OC}$= 0.67 V, $J_{SC}$= 43.29 mA.cm$^{-2}$, FF= 69.87 % and PCE of ~20.18% in a ITO/WS2/LNMO/CFTS/Au as a best solar structure after investigating several architectures and ETL/HTL materials using combined DFT calculation and SCAPS-1D device simulations [4]. *Abhishek Raj et al.* used SCAPS-1D to simulate several ETLs and obtained the a maximum PCE of ~ 0.18% with a $C_{60}$/Perovskite/CuI structure, this PSC however had poor PV performances ($J_{SC}$= 0.1982 mA/cm$^2$, $V_{OC}$= 1.9110V, and FF= 47.82%) [5]. The same group of *Abhishek Raj et al.* used SCAPS-1D in recent research and reported higher PV performances for a FTO/TiO2/LNMO/CuI/Au solar cell as $V_{OC}$= 0.764 V, $J_{SC}$=40.965 mA/cm$^2$, FF~ 85.6%, and PCE of ~26.8%[16]. These promising structures are somewhat hard to fabricate and require expensive/rare HTL materials, therefore, another innovative configuration could be explored. The HTL-free architecture has emerged recently as a cheaper, stable and efficient alternative to the common HTL/perovskite/ETL configuration [17]. The first HTL-free PSC has been fabricated by *Etgar et al.,* it consisted of a mesoscopic methylammonium lead iodide ($CH_3NH_3PbI_3$) perovskite/$TiO_2$ heterojunction [18]. This solar cell demonstrated impressive PV performance, with $J_{SC}$= 16.1 mA/cm$^2$, $V_{OC}$ = 0.631 V, FF = 57%, and a PCE of 5.5% under standard AM 1.5 solar light. This finding paved the wat for simpler, cheaper and efficient PV devices. Interest in HTL-free PV cells is growing, several research works are demonstrating excellent PV performances, *Eli Danladi et al.*, for example, used SCAPS-1D to investigate a Li-doped ETL/perovskite structure obtaining performance as PCE ~26.72%, FF of 85.53%, $J_{SC}$ of 22.265 mA/cm$^2$ and $V_{OC}$ = 1.387 V [19]. In this work, we suggest an inverted HTL-free SNMO-based PSC with an appropriate ETL and optimized architecture. For this purpose, the SCAPS-1D code is solicited. The paper is organized as follows, after the introduction, we introduce the SCAPS-1D software and its working principle, then we proceed with reproducing and experimental device with FTO/SNMO structure to establish a baseline. Subsequently, the inverted architecture (SNMO/ETL) is introduced, and the selection of the optimal ETL is achieved through a comprehensive systematic analysis. Further optimization explores the effects of absorber and ETL thickness, temperature, and series resistance on device performances. Simulations were performed under standard 1 Sun illumination at ambient temperature, demonstrating the potential of the proposed architecture for efficient and scalable lead-and-HTL-free photovoltaic devices.

## II. Simulation procedures

### II.1. SCAPS-1D

Our simulation work has been carried out using the Solar Cell Capacitance Simulator SCAPS-1D (3.3.11, 08-10-2024). Widely known and used by the PV research community [3, 4, 20-26], SCAPS-1D was developed by *Pr. Marc Burgelman.et.al* [27-30] at the Department of Electronics and Information Systems (ELIS) of the University of Gent (Belgium). This free one-dimensional simulator is capable of handling solar cells containing up to seven layers, and the simulation can be managed to study the effects of variating one or several parameters and also allows the study of the impact of defect levels within bulk and at the interfaces. The working principle of SCAPS is based on solving fundamental semiconductor equations, including the Continuity Equations for electrons and holes with the approximate boundary conditions (Eq.1 and .2), and Poisson Equation for electrostatic potential used for semiconductors (Eq.3) in the one-dimensional approach assuming electron and hole densities do not change with time [4, 14, 22, 31]. The other fundamental equations also treated within SCAPS-1D are the electron and hole charge transport Equations (4) and (5), respectively, and the absorption coefficient $\alpha(\lambda)$ (6), which are solved until convergence occurs [32].

$$\frac{d^2}{dx^2}\varphi(x) - \frac{q}{\varepsilon\varepsilon_0}\left(p(x) - n(x) + N_D - N_A + \rho_p - \rho_n\right) = 0 \qquad (1)$$

$$-\left(\frac{1}{q}\right)\frac{\partial J_p}{\partial x} + G(x) - R(x) = 0 \qquad (2)$$

$$-\left(\frac{1}{q}\right)\frac{\partial J_n}{\partial x} + G(x) - R(x) = 0 \qquad (3)$$

$$J_n = D_n \frac{\partial n}{\partial x} + \mu_n n \frac{\varphi(x)}{\partial x} \qquad (4)$$

$$J_p = D_p \frac{\partial p}{\partial x} + \mu_p p \frac{\varphi(x)}{\partial x} \qquad (5)$$

$$\alpha(\lambda) = \left(A + \frac{B}{h\upsilon}\right)\sqrt{h\upsilon - E_g} \qquad (6)$$

Where:

- $\varphi(x)$ is the electrostatic potential within the perovskite absorber.
- $q$ is the elementary charge.
- $\varepsilon_0$ is the permittivity of vacuum.
- $\varepsilon$ is the perovskite's relative permittivity.
- $N_A$ and $N_D$ present the acceptor and donor concentration, respectively.
- Concentration of electrons (holes) is denoted by $n$ ($p$).
- $\rho_p$ ($\rho_n$) signifies the hole (electron) distribution function.
- $J_p(J_n)$ signifies the current densities of hole (electron).
- $G$ presents the particle photogeneration rate.
- $R$ denotes the total recombination resulting from both direct and indirect recombination processes.
- $J_n$ and $J_p$ are electron and hole current densities, respectively.
- A and B are constants.
- $h$ is the Planck constant.
- $v$ is the frequency of photons.
- $E_g$ is the band gap of the absorber layer.

Recombination in deep bulk levels and their occupation is described by means of the Shockley-Read-Hall formalism (SRH), while, recombination at the interface states is described by an extension of the SRH formalism [3].

### II.2. Material properties and parameters

In this paper, the lead-free Samarium-based double perovskite $Sm_2NiMnO_6$ (SNMO) is used as the active absorbing layer; it crystallizes in the centrosymmetric monoclinic space group $P2_1/n$ (No.14, $C^5_{2h}$) [6]. In the unit cell, Sm and O atoms are at 4e sites of general $C_1$ symmetry while Ni/Mn atoms occupy the 2a and 2b sites of $C_i$ symmetry, (Wyckoff notation) [8, 9, 33]. In order to evaluate the photocurrent response, *Sheikh M.S. et al.,* prepared an FTO/SNMO/Au structure and exposed it to a 1 Sunlight intensity and proved a relatively high photocurrent density compared to $BiFeO_3$ or $KBiFe_2O_5$-based devices [8]. We found that this "HTL-free" n-i configuration (FTO/SNMO/Au) presents a working PV device meriting further investigation, which will be demonstrated in this work.

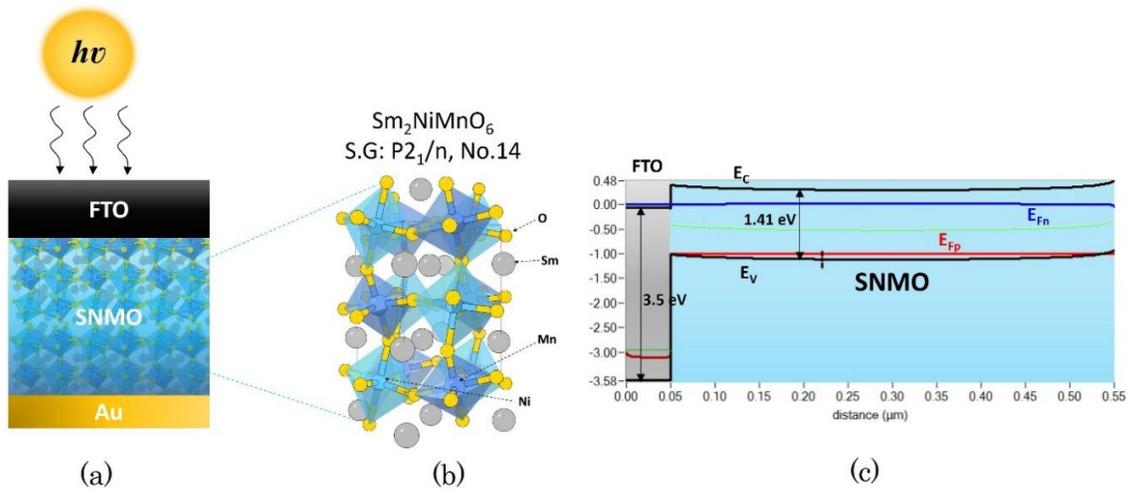

**Figure 1.** (a) Structure of the SNMO n-i solar cell, (b) crystalline structure of SNMO and (c) energy band diagram of the FTO/SNMO/Au PSC.

**Table 1:** Input parameters for the SCAPS-1D model in the i-n (SNMO/FTO) configuration

| Parameter | FTO [5, 19] | SNMO [6, 8, 9] | Li-TiO$_2$ [19] | TiO$_2$ [5] |
|---|---|---|---|---|
| Role | ETL | Absorber | ETL | ETL |
| Thickness w(nm) | 50 | 500 | 50 | 50 |
| Bandgap energy $E_g$ (eV) | 3.5 | 1.41 | 3 | 3.2 |
| Electron affinity $\chi$ (eV) | 4 | 3.52 | 4.2 | 3.9 |
| Relative dielectric permittivity $\varepsilon_r$ | 9 | 120 | 13.6 | 32 |
| Electron mobility $\mu_n$ (cm$^2$/Vs) | 20 | 22 | 2.0E-4 | 20 |
| Hole mobility $\mu_p$ (cm$^2$/Vs) | 10 | 22 | 3.0E-6 | 10 |
| CB Effective density of states $N_C$ (cm$^{-3}$) | 2.2E18 | 1.0E18 | 3.0E18 | 1.0E19 |
| VB Effective density of states $N_V$ (cm$^{-3}$) | 1.8E19 | 1.0E18 | 2.0E19 | 1.0E19 |
| Electron thermal velocity $v_e$ (cm/s) | 1.0E7 | 1.0E7 | 1.0E7 | 1.0E7 |
| Hole thermal velocity $v_h$ (cm/s) | 1.0E7 | 1.0E7 | 1.0E7 | 1.0E7 |
| Acceptor density $N_A$ (cm$^{-3}$) | 0 | 1.0E15 | 0 | 0 |
| Donor density $N_D$ (cm$^{-3}$) | 1.0E19 | 1.0E15 | 1.0E19 | 1.0E17 |
| Defects density $N_t$ (cm$^{-3}$) | 1.0E14 | 1.0E14 | 1.0E14 | 1.0E14 |

**Table 2:** Defects parameters of the SNMO perovskite layer used in the simulation

| Parameter | Defect 1 (SNMO) |
|---|---|
| Charge type | Neutral |
| Total density (cm$^{-3}$) | 1.0E18 |
| Energetic distribution | Gaussian |
| Energy level (eV) | 0.60 (above $E_V$) |
| Characteristic energy (eV) | 0.10 |
| Capture cross section (cm²) | 1.0E-15 |
| Capture cross section holes (cm²) | 1.0E-15 |
| $N_t$ total (cm$^{-3}$) | 1.0E14 |

**Table 3:** Defects parameters of the n-FTO/i-SNMO interface.

| Parameter | FTO/SNMO |
|---|---|
| Defect type | Single acceptor (0/−) |
| Energetic distribution | Gaussian |
| Capture cross section (cm²) | 1.0E-15 |
| Capture cross section holes (cm²) | 1.0E-15 |
| Reference for defect energy level $E_t$ | Above highest $E_v$ |
| Total density ($N_t$) (1/cm²) | 2.0E14 |
| Energy with respect to Reference (eV) | 0.6 |
| Characteristic energy (eV) | 0.1 |

## III. Results and discussion

We started the simulation work by reproducing an experimental model consisting of FTO as an electron transport layer (ETL) and SNMO as the perovskite absorber, while the back metallic contact is gold (work function =5.3eV). Initial simulation parameters for the simulated device are collected from theoretical and experimental literature and are listed in Table 1. The starting values of thicknesses, not found in literature, are inspired by similar solar cells in the n-i-p configuration and optimized in what follows. Values of defects in the materials and their interface (ETL/SNMO) are summarized in Table 2 and optimized hereafter. Defects of the ETL were taken as neutral, and Gaussian energetic distribution is adopted with a characteristic energy of 0.1 eV with a trap defect density of $1\times10^{12}$ cm$^{-2}$. The results of the first simulation step are illustrated through the J-V characteristic shown in Figure 02(a). We obtained a PCE of 7.81%, a fill factor (FF) of ~80.18%, an open-circuit voltage ($V_{OC}$) of ~0.8V, and a short-circuit current density ($J_{SC}$) of ~12.2 mA/cm2. The $J_{SC}$ is expected to be high (in the mA range), as experimentally reported in[8]. The simulated external quantum efficiency EQE of the solar cell, which reflects a ratio of the collected charge carriers by the solar cell to the total number of incident photons at a specified wavelength, is illustrated in Figure 02(b). It can be seen that this device configuration shows a narrow spectrum where EQE exceeds 40% from 450nm to 680nm, reaching a maximum of ~82% at 585nm. Next, we introduced a very thin layer of TiO$_2$ (5-15nm) between the SNMO absorber and the FTO layers, this keeps the solar cell HTL-free while likely enhancing its efficiency. TiO$_2$ plays an important role as a buffer layer having an intermediate bandgap ($E_{g(TiO2)}$=3.2 eV [5]) between the $E_{g(FTO)}$ =3.5 eV [5] and $E_{g(SNMO)}$=1.41 eV [8]. Adding a TiO$_2$ ETL layer with thickness of 5 nm decreased $V_{OC}$ from 0.8V to ~0.75V, $J_{SC}$ remained almost the same at ~12.63mA/cm$^2$, FF as well remained unchanged while PCE slightly increased from 7.81% to ~8.14%. The transparent fluorine-doped tin FTO is an efficient ETL but with the added TiO$_2$, which acts as an effective HTL as well as a "buffer layer", the cell performs better. This particular configuration is not of interest in this paper but will be soon investigated in a separate work.

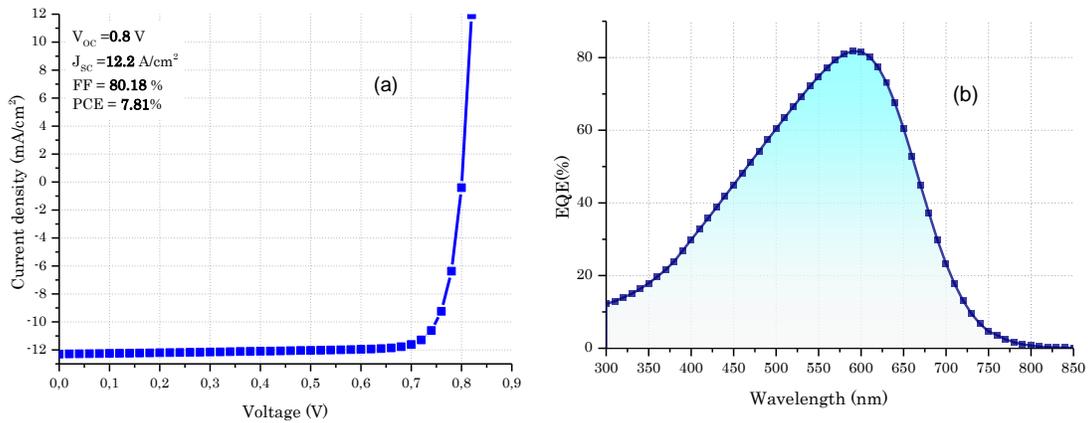

**Figure 2**. (a) J-V characteristic and (b) EQE vs wavelength of the initial n-i cell with FTO/SNMO/Au structure under 1Sun illumination.

### III.1. Inverted HTL-free SNMO PSC

In this step, we simulated an inverted version of our SNMO PSC. Inverted HTL-free PSCs are promising configurations offering cheaper alternatives to normal PSCs (ETL/Perovskite/HTL) by lowering production costs (as HTLs are expensive) and reducing the complexity of the cells, making them more competitive and thus highly required, hence promoting the commercialization of simpler, efficient, stable and non-toxic PSCs [17]. The suggested inverted SNMO SC (Au/SNMO/FTO, see figure 03(a)) simulated under 1 Sun illumination exhibited increased PV performances as $V_{OC}$=0.8 V, $J_{SC}$=16.46 mA/cm$^2$, FF=82.14%, and PCE=10.93%. EQE for this cell, shown in figure 03(b), along with the normal HTL-free cell, shows clear improvement of the spectrum, broadening its absorption in the visible range to ~60% @300nm, ~77%@400nm and a peak of ~93% @585nm. In order to improve these values, several solutions can be explored i.e. optimizing thicknesses and doping concentrations, adding buffer layers, changing the front layer, here taken as gold, to more convenient material having less diffusion effects into the perovskite layer which reduces their stability.

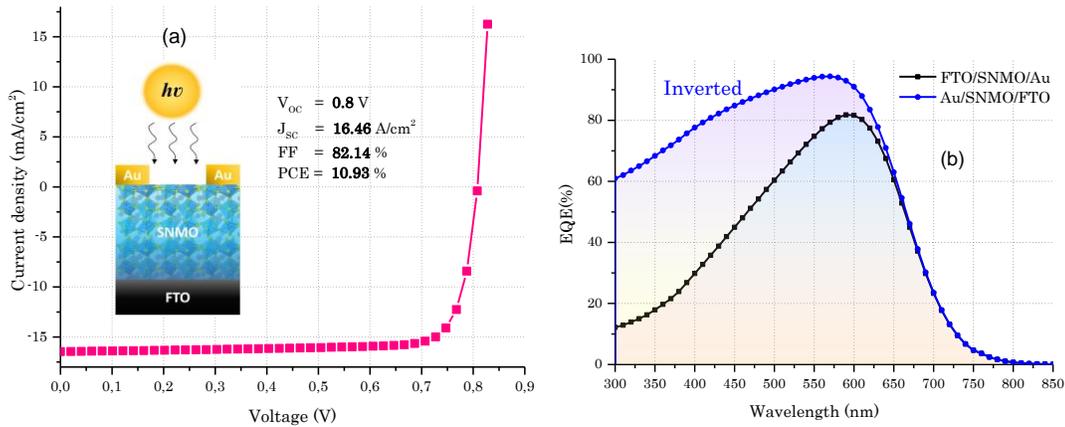

**Figure 03.** (a) J−V characteristic of the inverted Au/SNMO/FTO PSC showing its configuration and PV performance, (b) EQE of the inverted SC VS the normal FTO/SNMO/Au.

After showing the effectiveness of the inverted structure, we now simulate different available ETLs, keeping the thickness of SNMO at its initial value of 500nm, and that of the ETLs at 50 nm, this will be optimized in the next part. Simulation conditions are also kept the same. For this purpose, we choose several ETLs that are likely adapted to LnNiMnO$_6$ (Ln=La, Dy, Eu, Lu, and Sm) perovskites [4, 5, 15]. Among them are PCBM, ITO, ZnO, C$_{60}$, WS$_2$, TiO$_2$ and Li-TiO$_2$. Table 4 lists the main input parameters for the chosen ETLs as introduced to SCAPS-1D. HTL/SNMO interface defect parameters are kept the same as indicated in the Table3. Figure 04 illustrates the evolution of the PV performance through PCE, FF, V$_{OC}$ and J$_{SC}$ using different ETLs. It can be remarked that the best performances are recorded when the highly conductive indium tin oxide ITO is used (PCE~12.5%, FF~50.22, V$_{OC}$~1.507V, and J$_{SC}$~16.50mA/cm$^2$), this enhancement can be attributed to ITO's high doping concentration (N$_D$ = 1.0E21 cm$^{-3}$) and adequate electron/hole mobility ($\mu_e$ = 20 cm²V$^{-1}$s$^{-1}$, $\mu_h$ = 10 cm² V$^{-1}$s$^{-1}$). [4], this contributes to reducing the series resistance of the cell (Rs) and leads to an optimal conduction band offset with SNMO, as the ITO's conduction band minimum (CBM) lies at an energy level (~4 eV) that is well-matched with the conduction band of the SNMO absorber (3.52 eV), the energy alignment of adjacent layers enables efficient electron/hole extraction [4, 17]. Conversely, the lowest PV performances were recorded when TiO$_2$ and its Lithium-doped (Li-TiO$_2$) analog were adopted with PCEs of ~9.90% and ~7.42%, respectively; this is due to their higher dielectric constants and lower doping concentrations, which increase interface resistance and reduce charge collection efficiency. Moreover, Li-TiO$_2$ possesses extremely low charge carrier mobility, further degrading performance, which negatively affects electron/hole flow through the cell [5]. ZnO and PCBM demonstrated moderate performances, with PCEs of ~10.3% and ~10.6%, respectively. ZnO exhibited a balanced combination of conduction band alignment and electron mobility (100 cm²/V.s), while PCBM's lower electron mobility (0.2 cm²/V.s) was compensated by its efficient passivation of interface defects, as indicated by its high FF (~75.31%). WS$_2$ and C$_{60}$ delivered notable results, achieving PCEs of ~11.3% and ~10.9%, respectively. WS$_2$'s high doping concentration and favorable band alignment improved electron extraction, while $C_{60}$'s well-known strong light absorption in the visible region contributed to enhanced $J_{SC}$ values. However, both ETLs exhibited slightly lower FF compared to ITO, likely due to increased interface recombination [4].

**Table 4:** HTL Input parameters for the SCAPS-1D model in the i-n (SNMO/FTO) and p-i-n (NiO/SNMO/Nb:STO) configurations as compared with the experimental model in

| Parameter | ITO [4] | PCBM [4] | ZnO [4] | $C_{60}$ [4] | $WS_2$ [4] | $TiO_2$ [19] | Li-$TiO_2$ [19] |
|---|---|---|---|---|---|---|---|
| Thickness w(nm) | 500 | 50 | 50 | 50 | 100 | 30 | 30 |
| Bandgap energy $E_g$ (eV) | 3.5 | 2 | 3.3 | 1.7 | 1.8 | 3.2 | 3.0 |
| Electron affinity $\chi$ (eV) | 4 | 3.9 | 4 | 3.9 | 3.95 | 4.2 | 4.2 |
| Relative dielectric permittivity $\varepsilon_r$ | 9 | 3.9 | 9 | 4.2 | 13.6 | 10 | 13.6 |
| Electron mobility $\mu_n$ (cm²/Vs) | 20 | 0.2 | 100 | 8.0E-2 | 100 | 20 | 02E-4 |
| Hole mobility $\mu_p$ (cm²/Vs) | 10 | 0.2 | 25 | 3.5E-3 | 100 | 10 | 03E-6 |
| CB Effective density of states $N_C$ (cm⁻³) | 2.2E18 | 2.5E21 | 3.7E18 | 8.0E19 | 1.0E18 | 1.0E19 | 1.0E19 |
| VB Effective density of states $N_V$ (cm⁻³) | 1.8E19 | 2.5E21 | 1.8E19 | 8.0E19 | 2.4E19 | 1.0E19 | 1.0E19 |
| Acceptor density $N_A$ (cm⁻³) | 0 | 0 | 0 | 0 | 0 | 0 | 0 |
| Donor density $N_D$ (cm⁻³) | 1.0E21 | 2.93E17 | 1.0E18 | 1.0E17 | 1.0E18 | 1.0E17 | 1.0E19 |
| Defects density $N_t$ (cm⁻³) | 1.0E15 | 1.0E15 | 1.0E15 | 1.0E15 | 1.0E15 | 1.0E14 | 1.0E14 |

**Table 5:** comparison of PV parameters resulting from different ETLs introduced in the inverted HTL-free SNMO PSC

| HTL | $V_{OC}$ (V) | $J_{SC}$ (mA/cm²) | FF (%) | PCE (%) |
|---|---|---|---|---|
| FTO | 00.80 | 16.46 | 82.14 | 10.93 |
| $TiO_2$ | 04.23 | 16.41 | 12.90 | 09.90 |
| Li-$TiO_2$ | 01.04 | 16.20 | 44.14 | 07.42 |
| ITO | 01.51 | 16.50 | 50.22 | 12.50 |
| PCBM | 00.63 | 16.17 | 75.31 | 07.62 |
| $WS_2$ | 00.75 | 16.35 | 80.53 | 09.83 |
| $C_{60}$ | 00.71 | 16.14 | 72.68 | 08.28 |
| ZnO | 00.74 | 16.40 | 80.50 | 09.84 |

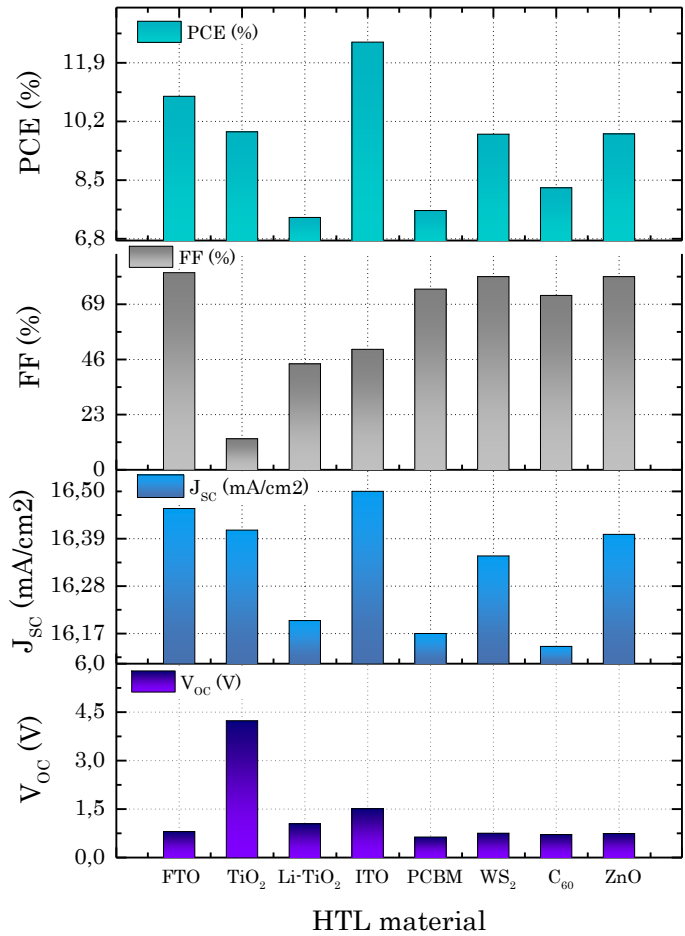

**Figure 04**. PV performance comparison of inverted HTL-Free SNMO PSCs using different ETLs.

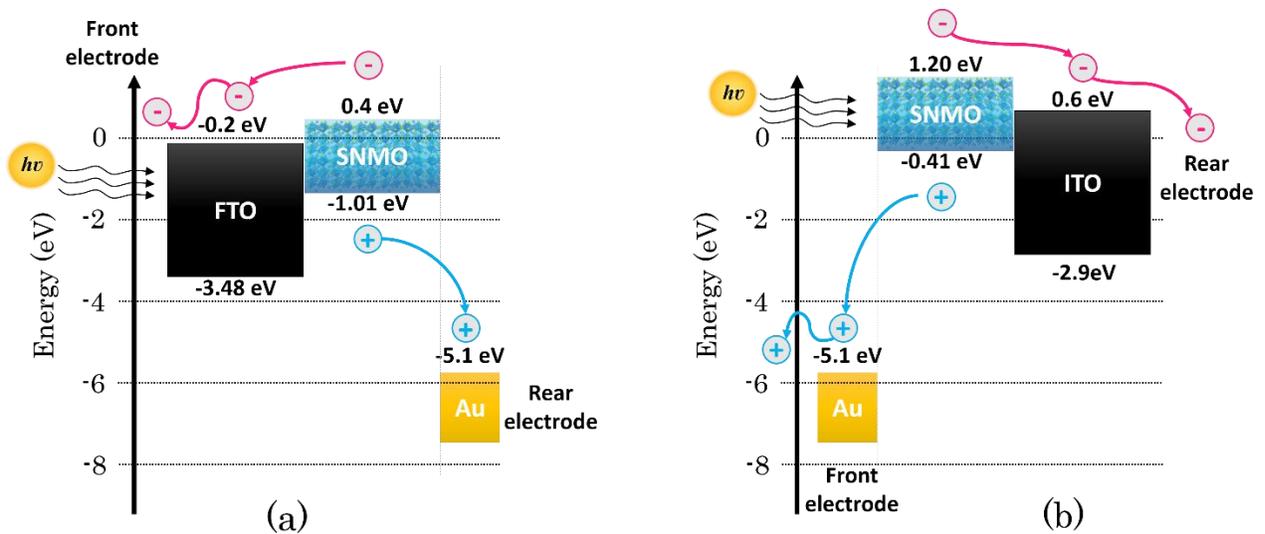

**Figure 05**. The energetic band diagram of (a) the initial FTO/SNMO/Au device, and (b) the suggested inverted and modified HTL-free Au/SNMO/ITO PSC.

To better understand the photovoltaic (PV) enhancement observed in the proposed inverted perovskite solar cell (PSC), we investigated the SCAPS-1D-estimated energetic band diagrams under 1 Sun illumination for both the initial FTO/SNMO device and the suggested SNMO/ITO inverted configuration, as shown in Figures 5(a) and 5(b), respectively. In the FTO/SNMO device (Figure 5(a)), the fluorine-doped tin oxide (FTO) front layer, with a wide bandgap ($E_g$=3.5 eV) and a valence band maximum (VBM) of -3.58 eV, exhibits poor alignment with the SNMO layer ($E_g$=1.41 eV, VBM ~ -1.0 eV). This mismatch creates a substantial valence band energy offset ($\Delta E_h$) of approximately 2.58 eV after band alignment, significantly impeding efficient hole transport. While the conduction band minimum (CBM) shows a smaller offset ($\Delta E_e$ ~ 0.6 eV), the overall imbalance between the valence and conduction band offsets leads to unbalanced carrier transport within the perovskite layer and at the FTO interface, substantially reducing PV performance. In contrast, the inverted SNMO/ITO configuration (Figure 5(b)) demonstrates a more favorable band alignment [17]. The ITO front electrode ($E_g$=3.5 eV, VBM ~ -2.9 eV) is better matched to SNMO, resulting in more balanced energy offsets on both the VBM and CBM sides. This improved alignment facilitates efficient hole extraction from the SNMO perovskite to the gold (Au) electrode and enables smooth electron transport through the ITO layer to the rear electrode. Further optimization could involve the use of materials with lower Fermi levels for the electrode or perovskites with higher VBMs [17]. Additionally, replacing the Au electrode with a transparent conductive material might enhance device versatility and transparency. A persistent challenge in PSCs is interfacial recombination, a dominant mechanism limiting device performance. This recombination is primarily driven by surface defects in contact layers, inadequate physical contact between layers, unbalanced carrier transport within the perovskite film, and energy mismatches between the perovskite and selective contacts. However, in high-quality perovskite films, bulk defect-mediated recombination can be minimized, shifting the focus to interface engineering as the primary determinant of performance [17]. Key interface-related issues such as electronic charge generation, extraction, injection, and recombination underscore the importance of valence band alignment at the perovskite/anode interface. For HTL-free inverted PSCs, the valence band energy offset ($\Delta E_h$) at this interface, defined by the VBM of the perovskite and the Fermi level of the anode, is particularly critical. Minimizing $\Delta E_h$ is essential for efficient charge injection and transport. Enhancing energy level alignment, either by lowering the Fermi level of ITO or increasing the VBM of the perovskite, represents a powerful strategy to overcome these barriers[17]. These findings highlight the intricate interplay between band alignment, interfacial recombination, and device performance, emphasizing the central role of interface engineering in the development of high-efficiency PSCs.

**III.2. Effect of absorber thickness**

Adopting the inverted configuration with ITO as an ETL, we proceed with optimizing the thickness of the SNMO perovskite layer. Exploiting the "batch" feature, one of the most powerful features of SCAPS-1D, we varied SNMO thickness from 50nm to 1000nm in 10 steps. Figure 06(a) shows the influence of SNMO thickness on the J-V characteristic; we notice the benefit of increasing SNMO thickness in enhancing the $V_{OC}$, However, its influence on $J_{SC}$ is minor. This can be attributed to reduced recombination losses in the absorber layer. A thicker SNMO (500nm to 1μm), enough to absorb efficiently enough light, according to the Beer-Lambert law and as the SNMO absorption coefficient is relatively higher ~$10^5 cm^{-1}$, and provides an enhanced charge separation due to its extended depletion region, leading to an improved built-in potential and reduced interface recombination, thus, higher $V_{OC}$ and then better PV performances. EQE is another measure to confirm this statement; in Figure 06(b), we notice a shift towards higher wavelengths when SNMO thickness is increased, and the peak ~95% moves from 380nm at 50nm to 620nm at 1µm. This indicates that a thicker absorber enhances light harvesting in the red region (by absorbing lower-energy photons) by increasing the optical path length and minimizing transmission losses. Figure 07 summarizes the impact of varying SNMO thickness on the main PV parameters. On different scales, each one of these parameters is positively influenced by thicker SNMO. The experimental limits are to be explored, and according to the synthesis and deposition technique, the quality, crystallinity, and enhanced contacts are to be carefully taken care of. We also tried to simulate the influence of the thickness of the ITO ETL; the results showed a minor impact on the main PV parameters; when ITO thickness was varied from 50nm to 100nm, PCE barely moved from ~12.5% to ~12.59%, FF from ~83.38% to ~83.4%. ITO thickness therefore has no major impact but it can be increased further if experimentally needed.

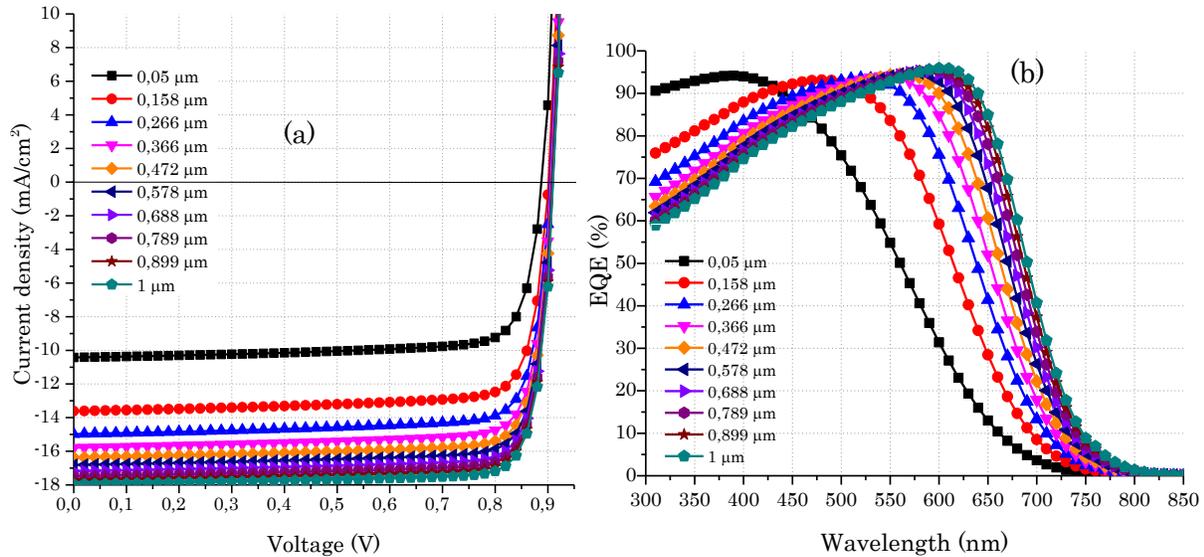

**Figure 06.** The evolution of (a) J-V characteristic and (b) the EQE of the SNMO/ITO solar cell with variation of SNMO thickness.

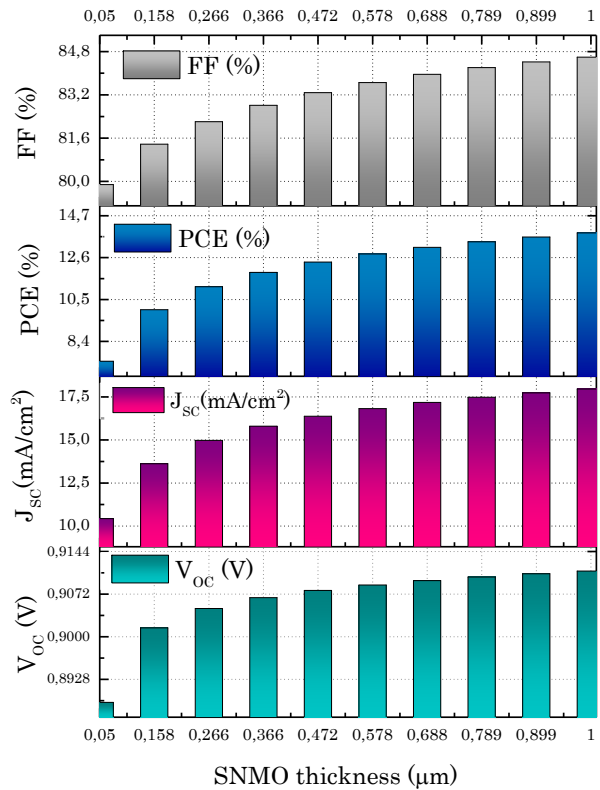

**Figure 07.** Impact of thickness of the SNMO layer on the main PV parameters of the studied solar cell.

### III.3. Effect of temperature and series resistance

The influence of temperature (Figure 08) and series resistance $R_S$ (Figure 09) on the PV performances of the inverted HTL-free SNMO solar cell are systematically analyzed. Increasing temperature from 250 K to 350 K showed a slight decline in $V_{OC}$ and enhancement in the fill factor, indicating enhanced thermal excitation and recombination at higher temperatures, which slightly decreases PCE. However, $J_{SC}$ remained largely unaffected. It can be said that the cell works well in the temperature range; its simple configuration and adequate material properties are the main reasons, but this needs to be confirmed experimentally. On the other hand, higher $R_S$ values, as seen in Figure 08, were found to significantly degrade device performances by lowering $J_{SC}$ and FF due to increased resistive losses. Surprisingly, when $R_S$ reaches ~500 $\Omega.cm^2$ and above, we noticed a strange increase in FF, while PCE, $V_{OC}$, and $J_{SC}$ slightly increased but at a very low rate. This highlights the need to minimize $R_S$ through improved contact quality and optimized material interfaces to achieve maximum efficiency in SNMO-based devices.

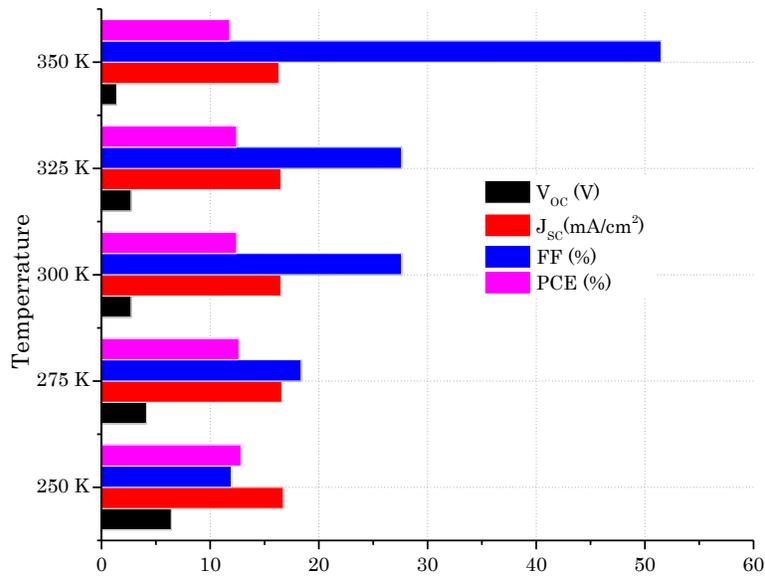

**Figure 08.** Influence of varying the temperature on the main PV parameters of the SNMO/ITO solar cell.

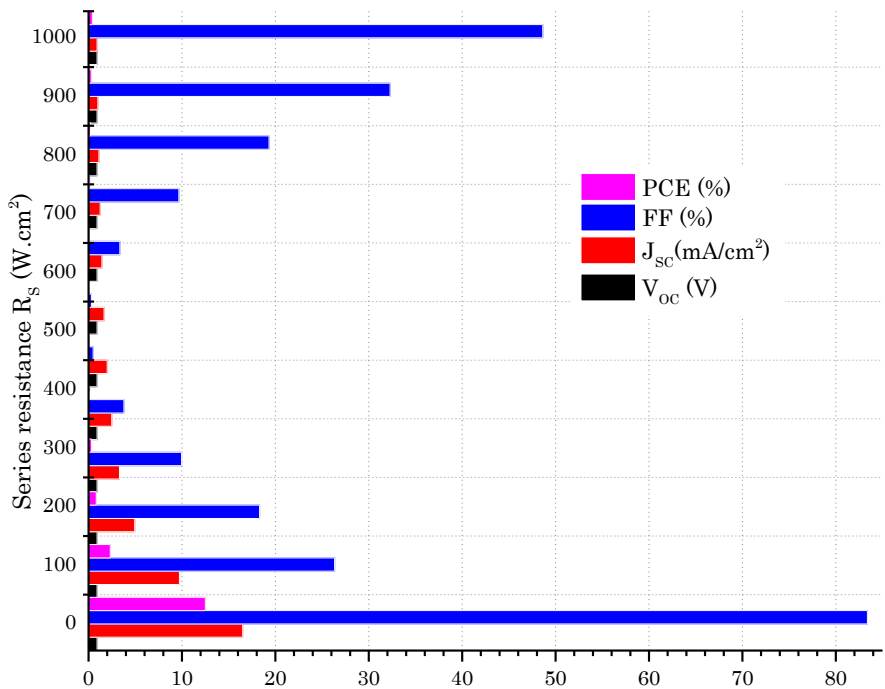

**Figure 09.** Influence of the series resistance $R_S$ on the main PV performances of the SNMO/ITO solar cell.

## IV. Conclusions

In this study, we solicited a SCAPS-1D simulator to explore the PV performances of HTL-free $Sm_2NiMnO_6$ (SNMO)-based perovskite solar cells, focusing on both standard and inverted configurations. Initial results demonstrated a PCE of ~7.81% for the standard configuration (FTO/SNMO/Au) and a significantly improved PCE of ~10.93% for the inverted structure (Au/SNMO/FTO). The inverted configuration exhibited enhanced performance metrics, including a higher $J_{SC}$=16.46 mA/cm², $V_{OC}$=0.8 V, FF of ~82.14%, and broader EQE absorption across the visible spectrum. Further analysis revealed that increasing the SNMO absorber thickness from 50 nm to 1 μm reduced recombination losses and extended light harvesting in the red region, shifting the EQE peak of about 96% from 380 nm to 620 nm. For 1um thickness of SNMO, PCE has been boosted to ~ 12.6%, FF to 84.6%, while $V_{CO}$ and $J_{SC}$ reached 0.91 V and 17.65mA/cm², respectively. These findings highlight the potential of SNMO as an efficient, stable, and lead-free absorber material for cheap HTL-free perovskite solar cells. However, there remains room for optimization, such as refining ETL choices, improving material crystallinity, and minimizing recombination losses. This work demonstrates the feasibility of integrating SNMO into next-generation photovoltaic devices, paving the way for sustainable and cost-effective solar energy solutions.